%% file: FabricEff.tex
\documentclass{llncs}

\usepackage{graphicx}
\usepackage{amsmath}
\usepackage{multirow}
\usepackage{listings}
\usepackage{xcolor}
\usepackage{algorithm}
\usepackage{algpseudocode}
\usepackage{pgfplots}
\pgfplotsset{compat=1.18}
\usepackage{caption}
\usepackage{subcaption}
\usepackage{tikz}
\usetikzlibrary{arrows.meta, positioning}
\usepackage{pgf-umlsd}
\usepackage{todonotes}

\usepackage{xspace}

\usepackage{tikz}
\usepackage{pgfplots}
\usepackage{subcaption}
\pgfplotsset{compat=1.18}
\input{macros}

\title{Dependency-Aware Execution Mechanism in Hyperledger Fabric Architecture}
\author{Sanyam Kaul \and
        Manaswini Piduguralla \and
        Gayathri Shreeya Patnala \and
        Sathya Peri}

\authorrunning{Kaul et al.}
\institute{Indian Institute of Technology Hyderabad, Telangana, India}

\begin{document}

\maketitle

\begin{abstract}
\label{sec:abs}
Hyperledger Fabric is a leading permissioned blockchain framework for enterprise use, known for its modular design and privacy features. While it strongly supports configurable consensus and access control, Fabric can face challenges in achieving high transaction throughput and low rejection rates under heavy workloads. These performance limitations are often attributed to endorsement, ordering, and validation bottlenecks. Further, optimistic concurrency control and deferred validation in Fabric may lead to resource inefficiencies and contention, as conflicting transactions are identified only during the commit phase.

To address these challenges, we propose a dependency-aware execution model for Hyperledger Fabric. Our approach includes: (a) a dependency flagging system during endorsement, marking transactions as independent or dependent using a hashmap; (b) an optimized block construction in the ordering service that prioritizes independent transactions; (c) the incorporation of a Directed Acyclic Graph (DAG) within each block to represent dependencies; and (d) parallel execution of independent transactions at the committer, with dependent transactions processed according to DAG order.

Incorporated in Hyperledger Fabric v2.5, our framework was tested on workloads with varying dependency levels and system loads. Results show up to 40\% higher throughput and significantly reduced rejection rates in high-contention scenarios. This demonstrates that dependency-aware scheduling and DAG-based execution can substantially enhance Fabric’s scalability while remaining compatible with its existing consensus and smart contract layers.
\end{abstract}


\section{Introduction}
\label{sec:intro}
\input{intro}

\section{Problem Statement and Contributions}
\label{sec:problem}
\input{pblm}

\section{Proposed Solution}
\label{sec:proposed}
\input{soln}

\subsection{Proposed Architecture}
\label{sec:psa}
\input{sys_arch}


\section{Performance Analysis}
\label{sec:perf}
\input{setup}

\label{sec:test}
\input{perf}

\section{Conclusion and Future Work}
\label{sec:conc}
\input{conc}

\bibliographystyle{splncs04}
\bibliography{citation_long}

\newpage
\appendix
\section{Appendix}

\input{append}

\end{document}

%% file: macros.tex
\newcommand{\cmnt}[1]{}
\newcommand{\ignore}[1]{}
\newcommand{\remove}[1]{}

\newcommand{\secref}[1]{Section~\ref{sec:#1}}
\newcommand{\figref}[1]{Fig.~\ref{fig:#1}}

\newcommand{\Secref}[1]{Section~\ref{sec:#1}}
\newcommand{\Figref}[1]{Figure~\ref{fig:#1}}
\newcommand{\Tabref}[1]{Table~\ref{tab:#1}}



%% file: intro.tex
Hyperledger Fabric is a permissioned blockchain platform widely used in enterprise applications~\cite{androulaki+:2018:eurosys}. Its modular design, with pluggable consensus and privacy features, enables industrial-grade flexibility. Fabric operates via a three-phase transaction pipeline: endorsement, ordering, and validation. Clients submit proposals to endorsers, who simulate transactions and return read-write sets. The ordering service collects these into blocks and broadcasts them to committing peers, who validate and apply them. Depending on its configuration, a peer node in the network can perform one or more of these roles: \emph{endorser}, \emph{orderer}, or \emph{committer}.

\begin{figure}[h]
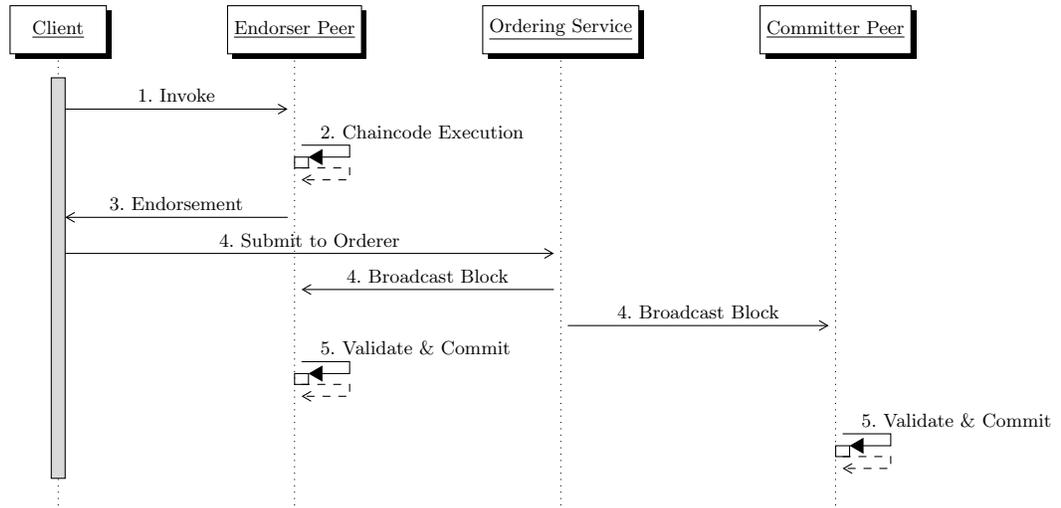

	\centering
	\tikzset{every picture/.append style={transform shape,scale=0.8}}
	\begin{sequencediagram}
		\newthread{client}{Client}
		\newinst[2]{endorser}{Endorser Peer}
		\newinst[2]{orderer}{Ordering Service}
		\newinst[2]{committer}{Committer Peer}
		
		\mess{client}{1. Invoke}{endorser}
		
		\begin{call}{endorser}{2. Chaincode Execution}{endorser}{}
		\end{call}
		
		\mess{endorser}{3. Endorsement}{client}
		
		\mess{client}{4. Submit to Orderer}{orderer}
		
		\mess{orderer}{4. Broadcast Block}{endorser}
		\mess{orderer}{4. Broadcast Block}{committer}
		
		\begin{call}{endorser}{5. Validate \& Commit}{endorser}{}
		\end{call}
		\begin{call}{committer}{5. Validate \& Commit}{committer}{}
		\end{call}
	\end{sequencediagram}
	\caption{Fabric transaction flow by Androulaki et.al ~\cite{androulaki+:2018:eurosys}}
	\label{fig:fabric_txn_flow}
\end{figure}

\vspace{1mm}
\noindent
\textbf{Hyperledger Fabric Architecture: }The transaction flow in Hyperledger Fabric, as illustrated in \figref{fabric_txn_flow}, begins when the client (\textbf{Step 1}) sends a transaction proposal to one or more \emph{endorser} peers. Each endorser peer then simulates the transaction by executing the specified chaincode (smart contract code, \textbf{Step 2}), generating a read/write set and an endorsement signature. The endorser returns this endorsement to the client (\textbf{Step 3}), and the client collects enough endorsements as required by the policy. The client then submits the endorsed transaction to the \emph{ordering service} (\textbf{Step 4}), which sequences transactions and packages them into blocks. The ordering service broadcasts the new block to all relevant peers, including both endorsers and \emph{committers} (\textbf{Step 4}). Finally, each committer peer validates the transactions in the block, checking endorsement policies and for conflicts, and commits valid transactions to the ledger (\textbf{Step 5}).

While this architecture performs well in low-contention environments, it exhibits significant limitations under high transaction volume. A key bottleneck is the lack of early dependency detection and parallelism during the commit phase. Fabric employs an optimistic concurrency model with versioning, where all transactions are allowed to proceed until final validation. Conflicts are detected only during the commit stage, based on world-state version checks. As a result, conflicting transactions are often rejected later in the commit stage, leading to increased retries and wasted compute resources.

For example, consider 100 transactions that want to reduce the same asset by a value of 100. Suppose the initial value of the asset of also 100. In such a scenario, only the first transaction commits successfully and updates the state to zero, while the remaining 99 are rejected due to stale versioning. Despite being endorsed and ordered, these transactions fail at the commit stage (\textbf{Step 5}), resulting in unnecessary latency and reduced throughput. In this paper, we wish to address this shortcoming. 

Another constraint is Fabric’s strictly sequential commit logic, which executes all transactions in a block one after another—even if many are independent. This underutilizes multicore systems and limits concurrency, making Fabric inefficient in high-dependency workloads.

%% file: pblm.tex
Hyperledger Fabric follows an \emph{execute-order-validate} model \cite{androulaki+:2018:eurosys} for processing transactions. Clients submit proposals that are simulated by endorsing peers. The results are passed to the ordering service, which sequences them into blocks. These blocks are then validated and committed by peers based on endorsement policies and version control. Although this pipeline is designed to decouple execution from ordering, it does not utilize parallelism effectively, especially at the commit stage.

The \textit{order-execute-validate} model is the traditional transaction processing paradigm adopted by early blockchains such as Bitcoin and Ethereum. In this approach, transactions are first collected and ordered into blocks through a consensus protocol. Once ordered, every node in the network sequentially executes all transactions in the agreed order, updating their local copy of the ledger state. Finally, nodes validate the results to ensure consistency across the network. This model enforces strict determinism in smart contract execution to guarantee that all nodes reach the same state. While conceptually simple, this approach imposes limitations on scalability and performance, as transactions must be executed sequentially and by all nodes. Furthermore, the requirement for deterministic code restricts the flexibility of smart contract development.

The \textit{execute-order-validate} model, pioneered by Hyperledger Fabric \cite{androulaki+:2018:eurosys}, reorders the transaction processing steps to address the limitations of the traditional approach. In this model, transaction proposals are first executed (or simulated) by a subset of peers, known as endorsers, which generate read/write sets and endorsements. These endorsed transactions are then ordered via a consensus service. After ordering, each peer validates the transactions against endorsement policies and checks for conflicts before committing them to the ledger. This separation of execution and ordering enables parallel transaction execution, improves throughput, and allows the use of general-purpose programming languages for smart contracts, as non-deterministic transactions can be detected and filtered out before ordering. The model is particularly suited for permissioned blockchains, where performance, flexibility, and strong consistency are essential.

A key issue lies in how Fabric handles transaction dependencies. Fabric employs an optimistic concurrency control mechanism with versioning, where all transactions are allowed to proceed to the commit phase regardless of conflicts. Final validation is deferred to the commit phase (stage 4), where transactions are checked for version consistency. If multiple transactions operate on the same key, only the one with the latest version is committed, while the rest are marked invalid. This leads to increased latency, unnecessary retries, and wasted compute resources for transactions that are eventually discarded.

For example (briefly discussed in \secref{intro}), consider an application using the asset-transfer chaincode where 100 concurrent transactions each attempt to deduct 100 units from the same asset, which has an initial balance of 100. Each transaction is endorsed based on the initial state and appears valid in isolation. However, during commit, only the first transaction updates the ledger successfully. The remaining 99 are rejected due to state version  mismatches, since their endorsements were based on stale data. This rejection is detected only in the commit phase and thus increasing the response time which could have been avoided had it been detected earlier in the endorsement phase. This illustrates the importance of conflict-aware application logic in Fabric and highlights how concurrent writes to the same key can degrade performance if not handled carefully.


Another major bottleneck is the strictly sequential commit logic. All transactions within a block are processed one after another, even if many are independent and could be executed concurrently. This prevents Fabric from leveraging multi-core architectures or exploiting transaction-level parallelism. The system lacks any mechanism to distinguish between independent and dependent transactions during execution.

These limitations result in two main issues:
\begin{enumerate}
	\item \textbf{High rejection rates under contention}: As shared state access increases, the likelihood of invalid transactions grows, leading to frequent rejections at the commit stage.
	\item \textbf{Under-utilization of system resources}: Independent transactions are serialized unnecessarily, limiting concurrency and throughput.
\end{enumerate}

These problems are particularly pronounced in enterprise applications, where multiple clients frequently interact with overlapping state variables. While previous enhancements to Fabric have targeted endorsement and ordering optimizations, they do not resolve the core inefficiencies in the commit phase. To support high-throughput workloads with mixed transactional dependencies, a fundamentally new execution model is required—one that introduces minimal architectural changes and remains compatible with Fabric’s existing endorsement policies and smart contract interfaces. 

In this paper, we address these issues. Specifically, we present a dependency-aware transaction execution mechanism in the Fabric's architecture. Our approach detailed in \Secref{proposed} includes: (a) a dependency flagging system during endorsement, marking transactions as independent or dependent using a hashmap; (b) an optimized block construction in the ordering service that prioritizes independent transactions; (c) the incorporation of a Directed Acyclic Graph (DAG) within each block to represent dependencies; and (d) parallel execution of independent transactions at the committer, with dependent transactions processed according to DAG order.

%% file: soln.tex
We introduce a transaction dependency-aware execution mechanism for Hyperledger Fabric that enhances parallelism and reduces transaction rejections. The objective is to detect dependencies early in the transaction lifecycle and use that information during commitment phase to reduce transaction rejections while enabling parallel execution wherever possible. These enhancements integrates cleanly within the existing Fabric architecture, requiring minimal disruption to its consensus or chaincode interfaces.

\begin{figure}[h]
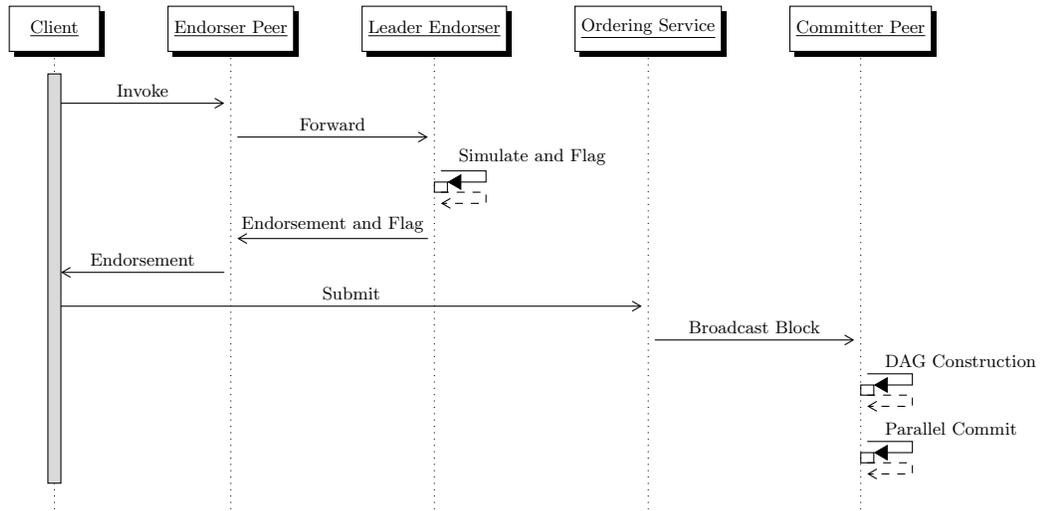

	\centering
	\tikzset{every picture/.append style={transform shape,scale=0.75}}
	\begin{sequencediagram}
		\newthread{client}{Client}
		\newinst[1.2]{endorser}{Endorser Peer}
		\newinst[1.2]{leader}{Leader Endorser}
		\newinst[1.2]{orderer}{Ordering Service}
		\newinst[1.2]{committer}{Committer Peer}
		
		\mess{client}{Invoke}{endorser}
		
		\mess{endorser}{Forward}{leader}
		
		\begin{call}{leader}{Simulate and Flag}{leader}{}
		\end{call}
		
		\mess{leader}{Endorsement and Flag}{endorser}
		
		\mess{endorser}{Endorsement}{client}
		
		\mess{client}{Submit}{orderer}
		
		\mess{orderer}{Broadcast Block}{committer}
		
		\begin{call}{committer}{DAG Construction}{committer}{}
		\end{call}
		\begin{call}{committer}{Parallel Commit}{committer}{}
		\end{call}
	\end{sequencediagram}
	\caption{Proposed Fabric transaction flow}
	\label{fig:proposed_fabric_leader_endorser}
\end{figure}

The proposed transaction flow for Hyperledger Fabric, as depicted in \figref{proposed_fabric_leader_endorser}, introduces a leader endorser and parallel execution mechanisms to enhance scalability and efficiency. The leader endorser coordinates the endorsement process across all the transaction received by various endorsing peers by maintaining the dependency relation. The process begins when the client sends a transaction proposal to an endorser peer, which then forwards the request to a designated leader endorser. The leader endorser simulates the transaction, flags any dependencies, and returns both the endorsement and dependency information to the original endorser. The endorser relays this endorsement back to the client, who submits the endorsed transaction to the ordering service. After ordering, the service broadcasts the block to the committer peer. The committer constructs a directed acyclic graph (DAG) based on the flagged dependencies and executes transactions in parallel where possible, committing valid results to the ledger. This architecture aims to maximize throughput by leveraging dependency-aware parallelism during the commit phase, while maintaining the integrity and consistency of the ledger state.

\subsection{Key Enhancements}

The solution introduces modifications across the endorsement, ordering, and commit phases, as described below:

\vspace{1em}
\noindent
\textbf{Dependency Flagging at Endorsement: }
We introduce a flagging mechanism at the leader endorser node. For each incoming transaction, we simulate its execution and inspect the key-value pairs it reads and writes. If a key has been recently modified or is part of an active endorsement set, the transaction is flagged as dependent (\texttt{flag = 1}); otherwise, it is marked as independent (\texttt{flag = 0}). The endorsement, along with the flag, is sent back to the client. Using this information, the client can decide whether to proceed with transaction submission or delay/abort it. By empowering the client with early conflict information, we reduce unnecessary network and validation overhead. This mechanism directly addresses the concurrency and commit-phase rejection issues detailed in \Secref{problem}.

A hashmap is maintained at the endorser to track recent key usage, logging transaction IDs and their accessed keys. Expiry timers are attached to purge stale metadata. This dependency flag is embedded into the transaction proposal response and is propagated through the pipeline.

For example, in a scenario where 100 transactions each attempt to reduce an asset’s value by 100 (starting from a value of 100), the first transaction—having no conflicting predecessors—is flagged as independent (\texttt{flag = 0}), while the rest are flagged as dependent (\texttt{flag = 1}) due to access on a stale world state.

\vspace{1em}
\noindent
\textbf{Propagation Through Ordering Phase: }
The ordering service remains mostly unchanged, maintaining compatibility with existing configurations (e.g., Raft, Kafka). However, it is modified to preserve and forward the dependency metadata. Custom fields in the block structure store transaction flags and associated key identifiers, ensuring the committer can reconstruct dependencies accurately without recomputation.

\vspace{1em}
\noindent
\textbf{DAG Construction and Parallel Execution at Commit:}
To enable fine-grained concurrency while preserving Fabric’s consistency guarantees, we enhance the committer to construct a lightweight Directed Acyclic Graph (DAG) from the transactions in each block. In this DAG, each transaction is represented as a node, and edges denote key-level dependencies. An edge from transaction A to B indicates that B depends on A’s outcome and must be executed afterward. A modified topological sort organizes transactions into DAG levels, where independent transactions (\texttt{flag = 0}) form the base level and can be executed in parallel, while dependent transactions (\texttt{flag = 1}) are layered above based on their dependency chains.

We update the commit worker logic to process transactions in DAG-level batches. Each level is handled using a thread pool, allowing concurrent execution of all transactions in that level. Once a level is completed, the next level is scheduled, respecting dependency constraints. If a transaction fails validation (e.g., due to a version mismatch), it is rejected, and its downstream transactions are re-evaluated with updated context. This ensures consistency while avoiding cascade failures. All concurrency is managed within the peer node—no client-side changes are needed.

\vspace{1em}
\noindent
\textbf{Endorsement Expiry and Conflict Cleanup:}
To prevent stale metadata from affecting performance, expiry timers are set on all endorsement flags and dependency entries. If a transaction remains uncommitted beyond a defined threshold, its dependency information is purged. This ensures the system remains responsive and avoids memory bloat. In such cases, the client must resubmit the transaction as a new request. Any endorser receiving a transaction with expired metadata will discard it.

%% file: sys_arch.tex
In this section we detail the modified architecture of Hyperledger Fabric as proposed in this work. The key idea is to integrate a dependency-aware execution mechanism while retaining the core stages of Fabric’s pipeline: endorsement, ordering, and commit. The modified setup introduces changes to each phase, focused on tracking dependencies, constructing a DAG from transaction metadata, and enabling level-wise parallel execution at the commit stage.

\vspace{1em}
\noindent
\textbf{Modified Endorsement Logic:} In the standard Fabric flow, endorsers simulate the transaction and return the read-write sets to the client without explicitly tracking dependencies. We extend the logic of the \textbf{leader endorser} to inspect whether the keys involved in the transaction have already been accessed by other active transactions.

A key-value hashmap is maintained to store active keys and the latest transaction IDs associated with them. The endorsement logic is modified as follows:

\paragraph{PCode: Leader Endorser}
\begin{algorithm}
	\caption{Transaction Endorsement with Dependency Flagging}
	\begin{algorithmic}[1]
		\Require Transaction $T$ received
		\State Simulate $T$
		\If{simulation fails}
		\State Reject $T$
		\State Notify client
		\Else
		\If{variable in $T$ exists in HASHMAP}
		\State $flag \gets 1$
		\State Add dependency reference to HASHMAP
		\Else
		\State $flag \gets 0$
		\State Add variable entry to HASHMAP
		\EndIf
		\State Set endorsement expiry timer
		\State Sign $T$ with $flag$ and dependency metadata
		\State Return to client
		\EndIf
	\end{algorithmic}
	\label{alg:leader_endorser}
\end{algorithm}

The pseudocode for the leader endorser's transaction endorsement with dependency flagging is outlined in Algorithm~\ref{alg:leader_endorser}. Upon receiving a transaction $T$ (line 1), the leader endorser first simulates its execution (line 2). If the simulation fails (line 3), the transaction is rejected and the client is notified (lines 4--5). If the simulation succeeds (line 6), the algorithm checks whether any variable accessed by $T$ already exists in the HASHMAP (line 7). If such a variable is found, a dependency flag is set ($flag \gets 1$), and a reference to this dependency is added to the HASHMAP (lines 8--9). Otherwise, the flag is set to zero and a new entry for the variable is created (lines 10--11). The endorser then sets an expiry timer for the endorsement (line 12), signs the transaction along with the flag and dependency metadata (line 13), and returns the result to the client (line 14). This process enables the system to efficiently track and signal transaction dependencies, facilitating subsequent parallel execution while preserving consistency.

This logic ensures that transactions accessing shared state are marked as dependent, and the information is embedded directly into the transaction metadata returned to the client.

\vspace{1em}
\noindent
\textbf{DAG-Based Block Construction and Commitment:} After transactions are ordered into blocks, they are forwarded to committing peers. The commit logic is extended to construct a lightweight Directed Acyclic Graph (DAG) using the metadata in the block.


\begin{algorithm}
	\caption{DAG-based Parallel Transaction Processing}
	\begin{algorithmic}[1]
		\Require Block received from ordering service
		\State Construct DAG:
		\ForAll{transaction $t$ in block}
		\If{flag$(t) == 0$}
		\State Add $t$ to Level 0 (independent)
		\Else
		\State Add edge from $t$ to parent transaction(s) based on dependency
		\EndIf
		\EndFor
		\State Process DAG:
		\ForAll{DAG level $L$}
		\ForAll{transaction $t$ in $L$ \textbf{in parallel}}
		\State Validate $t$
		\If{valid}
		\State Update world state with $t$
		\State Notify peers and ordering service
		\Else
		\State Reject $t$
		\EndIf
		\EndFor
		\EndFor
	\end{algorithmic}
	\label{alg:dag_parallel_processing}
\end{algorithm}

Algorithm~\ref{alg:dag_parallel_processing} describes the DAG-based parallel transaction processing approach used by the committer peer. Upon receiving a block from the ordering service (line 1), the committer constructs a directed acyclic graph (DAG) to capture transaction dependencies (line 2). For each transaction $t$ in the block (line 3), if the dependency flag is zero, $t$ is added to Level 0 of the DAG, indicating it is independent (lines 4--5). Otherwise, an edge is added from $t$ to its parent transaction(s) based on the recorded dependency information (lines 6--7). Once the DAG is constructed (line 9), transactions are processed level by level (line 10). Within each level, all transactions are validated and executed in parallel (lines 11--12). If a transaction is valid, the world state is updated and peers as well as the ordering service are notified (lines 13--15); invalid transactions are rejected (lines 16--17). This method enables safe parallelism by ensuring that dependent transactions are executed only after their prerequisites, thus improving throughput while maintaining correctness.

Transactions with \texttt{flag = 0} are processed in parallel immediately. Transactions with \texttt{flag = 1} are scheduled based on DAG dependencies. If two dependent transactions are at the same level and do not share keys, they can be executed in parallel. 

\begin{figure}[htbp]
	\centering
	\includegraphics[width=0.98\linewidth]{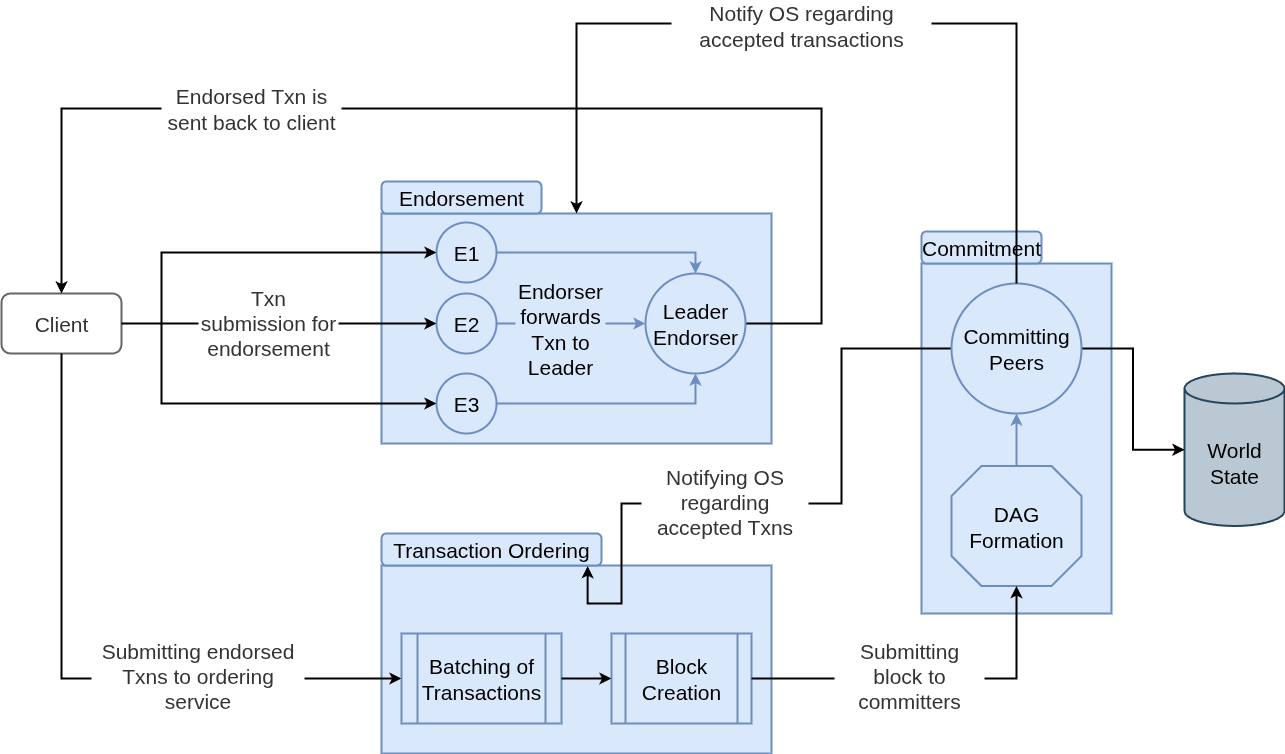}
	\caption{Proposed DAG-aware Fabric Architecture}
	\label{fig:modified-arch}
\end{figure}

\vspace{1em}
\noindent
\textbf{Architecture Overview:} Figure~\ref{fig:modified-arch} provides a high-level view of the modified architecture.

The architecture retains Fabric’s existing modular design. Clients continue to interact with the peer nodes through standard SDK interfaces. The only change is in how dependencies are internally detected, propagated, and processed. Endorsement flagging, DAG construction, and parallel validation are all encapsulated within the peer logic.

\vspace{1em}
\noindent
\textbf{Commit Thread Scheduling:} We integrate a thread pool at the committer that processes DAG levels concurrently. This thread pool dynamically scales with the number of level-0 and level-n transactions. The scheduler ensures that all parent transactions of a node are completed before the child transaction is validated.

The DAG is constructed using adjacency lists, and a topological sort is applied to determine the execution sequence. Transactions in the same level are passed to available threads in the pool.

\vspace{1em}
\noindent
\textbf{Compatibility and Modularity:} The proposed enhancements are backward-compatible with existing chaincode logic and network configurations. No changes are required in chaincode interface, world state structure, or ordering service APIs. This ensures that networks already using Fabric can adopt the new model with minimal transition effort.


\vspace{1em}
\noindent
\textbf{Expected Impact and Benefits:}The proposed enhancements deliver the following benefits over the default Fabric execution model:

\begin{itemize}
	\item \textbf{Lower rejection rates:} Early dependency detection reduces commit-time rejections, especially under high contention.
	\item \textbf{Better resource utilization:} Independent transactions are executed in parallel, improving CPU and memory usage.
	\item \textbf{Higher throughput and lower latency:} DAG-level concurrency significantly improves transaction throughput and reduces average response time (ART), as demonstrated by up to 40\% performance gain in our experiments.
	\item \textbf{Minimal architectural disruption:} The enhancements require only internal changes to peer logic; existing chaincode, clients, and ordering services remain untouched.
	\item \textbf{Scalability with workload dynamics:} The system adapts well to mixed workloads with varying dependency ratios.
\end{itemize}

\noindent
To evaluate the effectiveness of our proposed enhancements, we have carefully designed a series of experiments that simulate realistic and diverse blockchain workloads with varying levels of contention. These experiments are tailored to assess whether the expected benefits, such as reduced transaction rejection rates, improved resource utilization, and increased throughput, are achieved in practice. \Secref{perf} details the experiments design, the implementation platform, and the observations from results.

%% file: setup.tex
All experiments were conducted on a local development environment configured for reproducibility and performance isolation. The setup details are summarized in \Tabref{config}. \Tabref{components} and \Tabref{parameters} in the Appendix give a brief summary of the Fabric components modified by the authors, as well as the parameters set by the authors for workload simulation.

\begin{table}[h!]
	\centering
	\begin{tabular}{|l|l|}
		\hline
		\textbf{Component} & \textbf{Specification} \\
		\hline
		Processor & AMD Ryzen 5 5500U, 2.1GHz, 6 cores / 12 threads \\
		Memory & 14 GB DDR4 RAM \\
		Disk & 512 GB NVMe SSD \\
		Operating System & Ubuntu 24.04 LTS \\
		\hline
	\end{tabular}
	\caption{Hardware and Software Configuration}
    \label{tab:config}
\end{table}
\vspace{-2em}

%% file: perf.tex
\noindent
\textbf{Note:} All results presented here are based on the Voting Contract. Additional preliminary results for the Asset-Transfer Contract and Wallet Contract, are provided in the Appendix for further comparison and analysis.

\vspace{1mm}
\noindent
\textbf{Experiment Settings:} The experiments compare three threading strategies. (a) The first is the \emph{Original Fabric}, which serves as the baseline implementation without any concurrency optimizations. (b) The second approach, \emph{Modified Fabric (Dynamic Threads)}, employs a dynamic number of threads equal to the number of transactions at each DAG level, with the count capped by the number of physical cores to prevent oversubscription. (c) The third strategy, referred to as \emph{2-threaded / 4-threaded Fabric}, uses a fixed number of threads per DAG level, irrespective of the transaction count.

\label{sec:analysis}
\begin{figure}[htbp]
	\centering	
	\begin{subfigure}[b]{0.48\textwidth}
		\centering
		\begin{tikzpicture}
			\begin{axis}[
				xlabel={Number of Transactions},
				ylabel={Throughput (tx/sec)},
				legend style={at={(0.5,1.1)}, anchor=south, cells={align=left}, font=\small},
				legend columns=2,
				xtick={1000,2000,3000,4000,5000},
				ymajorgrids=true,
				grid=major,
				width=\textwidth,
				height=0.7\textwidth
				]
				\addplot coordinates {(1000,0.178) (2000,0.228) (3000,0.262) (4000,0.283) (5000,0.276)};
				\addplot coordinates {(1000,0.238) (2000,0.315) (3000,0.353) (4000,0.373) (5000,0.384)};
				\legend{Original Fabric, Modified Fabric}
			\end{axis}
		\end{tikzpicture}
		\caption{Original vs Modified Fabric}
	\end{subfigure}
	\hfill
	\begin{subfigure}[b]{0.48\textwidth}
		\centering
		\begin{tikzpicture}
			\begin{axis}[
				xlabel={Number of Transactions},
				ylabel={Throughput (tx/sec)},
				legend style={at={(0.5,1.1)}, anchor=south, cells={align=left}, font=\small},
				legend columns=2,
				xtick={1000,2000,3000,4000,5000},
				ymajorgrids=true,
				grid=major,
				width=\textwidth,
				height=0.7\textwidth
				]
				\addplot coordinates {(1000,0.178) (2000,0.226) (3000,0.261) (4000,0.283) (5000,0.276)};
				\addplot coordinates {(1000,0.234) (2000,0.316) (3000,0.352) (4000,0.371) (5000,0.384)};
				\addplot coordinates {(1000,0.228) (2000,0.290) (3000,0.349) (4000,0.356) (5000,0.336)};
				\addplot coordinates {(1000,0.236) (2000,0.316) (3000,0.307) (4000,0.336) (5000,0.335)};
				\legend{Original, Modified, Modified-2T, Modified-4T}
			\end{axis}
		\end{tikzpicture}
		\caption{Parallelism in Modified Fabric}
	\end{subfigure}
    \caption{Experiment 1: Impact of Number of Transaction on Throughput}
    \label{fig:exp1}
\end{figure}
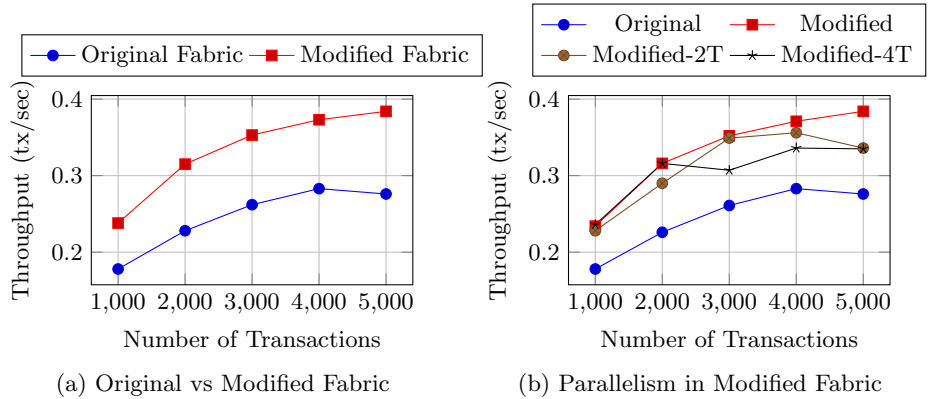

\noindent
\textbf{Experiment 1: Impact of Number of Transaction on Throughput} \\
This experiment illustrated in \Figref{exp1} evaluated the throughput across increasing transaction loads. The Modified Fabric (Dynamic Threads) consistently achieved higher throughput than the Original Fabric. At 5000 transactions, throughput improved from 0.276 tx/sec to 0.384 tx/sec (approx. 39\% gain). Introduction of fixed-thread variants showed limited or mixed improvements. While 2-threaded execution showed some benefit, the 4-threaded variant often plateaued or underperformed due to thread contention or underutilization depending on DAG width.

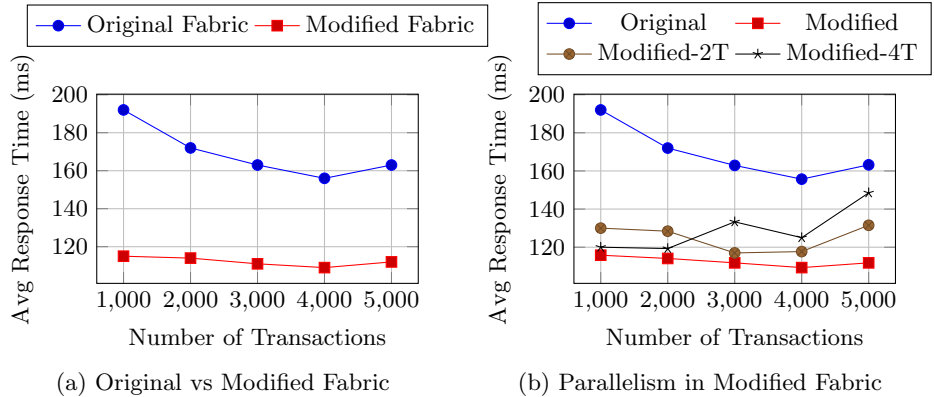
\begin{figure}[htbp]
	\centering
	\begin{subfigure}[b]{0.48\textwidth}
		\centering
		\begin{tikzpicture}
			\begin{axis}[
				xlabel={Number of Transactions},
				ylabel={Avg Response Time (ms)},
				legend style={at={(0.5,1.25)}, anchor=south, cells={align=left}, font=\small},
				legend columns=2,
				xtick={1000,2000,3000,4000,5000},
				ymajorgrids=true,
				grid=major,
				width=\textwidth,
				height=0.7\textwidth
				]
				\addplot coordinates {(1000,192) (2000,172) (3000,163) (4000,156) (5000,163)};
				\addplot coordinates {(1000,115) (2000,114) (3000,111) (4000,109) (5000,112)};
				\legend{Original Fabric, Modified Fabric}
			\end{axis}
		\end{tikzpicture}
		\caption{Original vs Modified Fabric}
	\end{subfigure}
	\hfill
	\begin{subfigure}[b]{0.48\textwidth}
		\centering
		\begin{tikzpicture}
			\begin{axis}[
				xlabel={Number of Transactions},
				ylabel={Avg Response Time (ms)},
				legend style={at={(0.5,1.1)}, anchor=south, cells={align=left}, font=\small},
				legend columns=2,
				xtick={1000,2000,3000,4000,5000},
				ymajorgrids=true,
				grid=major,
				width=\textwidth,
				height=0.7\textwidth
				]
				\addplot coordinates {(1000,192) (2000,172) (3000,162.9) (4000,155.7) (5000,163.2)};
				\addplot coordinates {(1000,115.8) (2000,114.1) (3000,111.8) (4000,109.3) (5000,111.8)};
				\addplot coordinates {(1000,130) (2000,128.4) (3000,117) (4000,117.7) (5000,131.5)};
				\addplot coordinates {(1000,120) (2000,119.3) (3000,133.3) (4000,125) (5000,148.5)};
				\legend{Original, Modified, Modified-2T, Modified-4T}
			\end{axis}
		\end{tikzpicture}
		\caption{Parallelism in Modified Fabric}
	\end{subfigure}
	\caption{Experiment 2: Latency Reduction Analysis}
    \label{fig:exp2}
\end{figure}

\vspace{1em}
\noindent
\textbf{Experiment 2: Latency Reduction Analysis} \\
This experiment (\Figref{exp2}) focused on average transaction latency. The Modified Fabric showed substantial latency reduction—for example, reducing latency from 192 ms to 115 ms at 1000 transactions. The 2-threaded and 4-threaded variants displayed varied performance. The 2-threaded variant showed modest improvements, while the 4-threaded version achieved lowest latency in some high-load cases, but exhibited inconsistent behavior due to rigid thread allocation at each DAG level.

\begin{figure}[htbp]
	\centering
	\begin{subfigure}[b]{0.48\textwidth}
		\centering
		\begin{tikzpicture}
			\begin{axis}[
				xlabel={Dependency Ratio},
				ylabel={Avg Response Time (ms)},
				legend style={at={(0.5,1.25)}, anchor=south, cells={align=left}, font=\small},
				legend columns=2,
				xtick={0,0.1,0.2,0.3,0.4,0.5,0.6,0.7,0.8,0.9},
				ymajorgrids=true,
				grid=major,
				width=\textwidth,
				height=0.7\textwidth
				]
				\addplot coordinates {(0,214) (0.1,224) (0.2,210) (0.3,205) (0.4,202) (0.5,192) (0.6,176) (0.7,175) (0.8,160) (0.9,152)};
				\addplot coordinates {(0,140) (0.1,139) (0.2,143) (0.3,137) (0.4,127) (0.5,115) (0.6,110) (0.7,102) (0.8,96) (0.9,88)};
				\legend{Original Fabric, Modified Fabric}
			\end{axis}
		\end{tikzpicture}
		\caption{Original vs Modified Fabric}
	\end{subfigure}
	\hfill
	\begin{subfigure}[b]{0.48\textwidth}
		\centering
		\begin{tikzpicture}
			\begin{axis}[
				xlabel={Dependency Ratio},
				ylabel={Avg Response Time (ms)},
				legend style={at={(0.5,1.1)}, anchor=south, cells={align=left}, font=\small},
				legend columns=2,
				xtick={0,0.1,0.2,0.3,0.4,0.5,0.6,0.7,0.8,0.9},
				ymajorgrids=true,
				grid=major,
				width=\textwidth,
				height=0.7\textwidth
				]
				\addplot coordinates {(0,215) (0.1,224) (0.2,210) (0.3,205) (0.4,202) (0.5,192) (0.6,176) (0.7,175) (0.8,160) (0.9,152)};
				\addplot coordinates {(0,140) (0.1,140) (0.2,144) (0.3,140) (0.4,128) (0.5,115) (0.6,110) (0.7,102) (0.8,95) (0.9,88)};
				\addplot coordinates {(0,163) (0.1,193) (0.2,142) (0.3,140) (0.4,135) (0.5,130) (0.6,122) (0.7,115) (0.8,110) (0.9,100)};
				\addplot coordinates {(0,140) (0.1,148) (0.2,134) (0.3,136) (0.4,130) (0.5,120) (0.6,119) (0.7,137) (0.8,108) (0.9,92)};
				\legend{Original, Modified, Modified-2T, Modified-4T}
			\end{axis}
		\end{tikzpicture}
		\caption{Parallelism in Modified Fabric}
	\end{subfigure}
		\caption{Experiment 3: Impact of Dependency Ratio on Latency}
        \label{fig:exp3}
\end{figure}
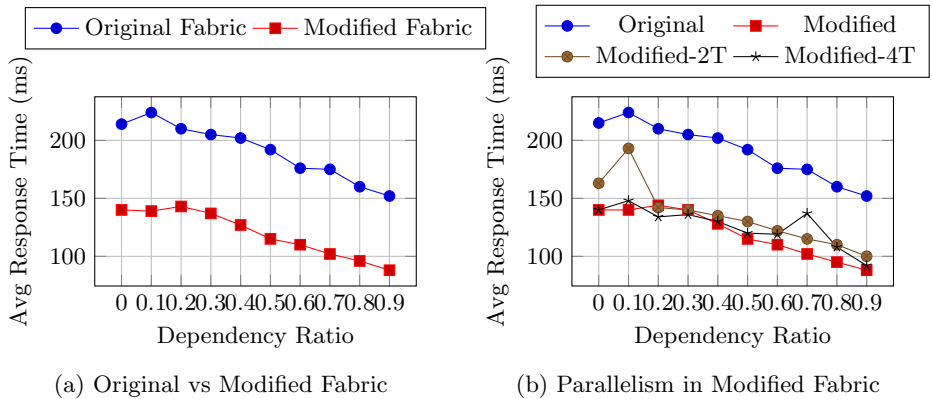

\vspace{1em}
\noindent
\textbf{Experiment 3: Impact of Dependency Ratio on Latency} \\
This experiment (\Figref{exp3}) analyzed performance under varying inter-transaction dependencies, measured as a dependency ratio from 0 to 0.9. As dependency increased, latency in the Original Fabric increased significantly, while the Modified Fabric with Dynamic Threads maintained much lower and more stable latencies. The 2-threaded and 4-threaded variants showed mid-range performance, with the 4-threaded approach performing best at high dependency levels (e.g., 92 ms at 0.9), though not always better than the dynamic strategy.

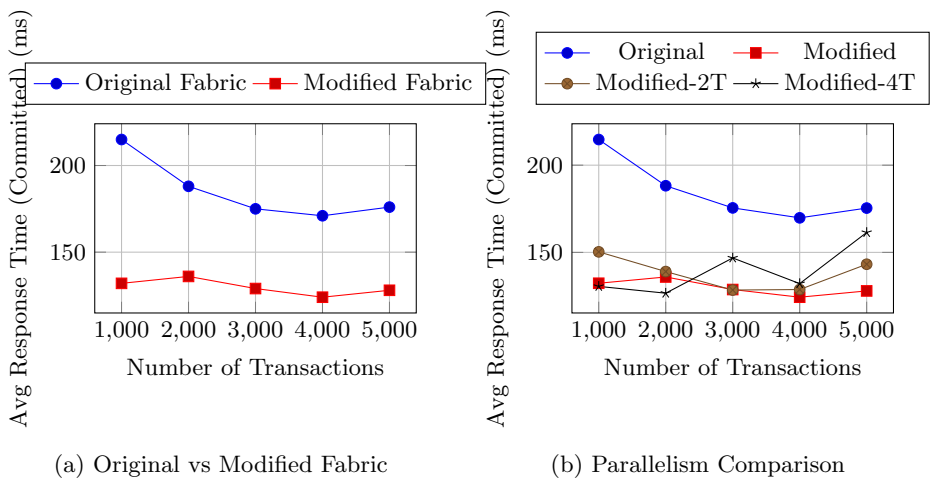
\begin{figure}[htbp]
	\centering
	\begin{subfigure}[b]{0.48\textwidth}
		\centering
		\begin{tikzpicture}
			\begin{axis}[
				xlabel={Number of Transactions},
				ylabel={Avg Response Time (Committed) (ms)},
				legend style={at={(0.5,1.1)}, anchor=south, font=\small},
				legend columns=2,
				xtick={1000,2000,3000,4000,5000},
				ymajorgrids=true,
				grid=major,
				width=\textwidth,
				height=0.7\textwidth
				]
				\addplot coordinates {(1000,215) (2000,188) (3000,175) (4000,171) (5000,176)};
				\addplot coordinates {(1000,132) (2000,136) (3000,129) (4000,124) (5000,128)};
				\legend{Original Fabric, Modified Fabric}
			\end{axis}
		\end{tikzpicture}
		\caption{Original vs Modified Fabric}
	\end{subfigure}
	\hfill
	\begin{subfigure}[b]{0.48\textwidth}
		\centering
		\begin{tikzpicture}
			\begin{axis}[
				xlabel={Number of Transactions},
				ylabel={Avg Response Time (Committed) (ms)},
				legend style={at={(0.5,1.1)}, anchor=south, font=\small},
				legend columns=2,
				xtick={1000,2000,3000,4000,5000},
				ymajorgrids=true,
				grid=major,
				width=\textwidth,
				height=0.7\textwidth
				]
				\addplot coordinates {(1000,214.7) (2000,188.2) (3000,175.5) (4000,169.8) (5000,175.4)};
				\addplot coordinates {(1000,132.3) (2000,135.9) (3000,128.7) (4000,124.3) (5000,127.9)};
				\addplot coordinates {(1000,150.3) (2000,139) (3000,128.4) (4000,128.7) (5000,143.2)};
				\addplot coordinates {(1000,130.5) (2000,126.6) (3000,146.8) (4000,132.1) (5000,161.4)};
				\legend{Original, Modified, Modified-2T, Modified-4T}
			\end{axis}
		\end{tikzpicture}
		\caption{Parallelism Comparison}
	\end{subfigure}
    \caption{Committed Transactions Results}
    \label{fig:committed}
\end{figure}

In addition to the core experiments, we evaluated the system performance by analyzing average response times for both committed and aborted transactions. These metrics provide further insight into the user-perceived latency and system efficiency.

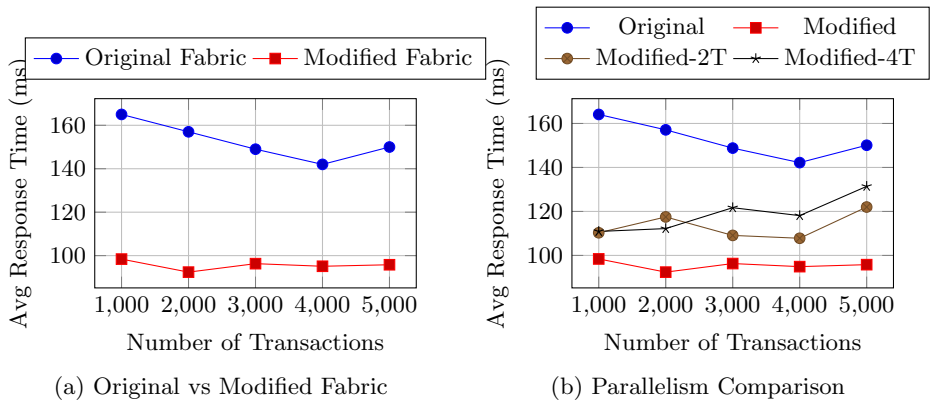
\begin{figure}[htbp]
	\centering
	\begin{subfigure}[b]{0.48\textwidth}
		\centering
		\begin{tikzpicture}
			\begin{axis}[
				xlabel={Number of Transactions},
				ylabel={Avg Response Time (ms)},
				legend style={at={(0.5,1.1)}, anchor=south, font=\small},
				legend columns=2,
				xtick={1000,2000,3000,4000,5000},
				ymajorgrids=true,
				grid=major,
				width=\textwidth,
				height=0.7\textwidth
				]
				\addplot coordinates {(1000,165) (2000,157) (3000,149) (4000,142) (5000,150)};
				\addplot coordinates {(1000,98.4) (2000,92.4) (3000,96.3) (4000,95.1) (5000,95.8)};
				\legend{Original Fabric, Modified Fabric}
			\end{axis}
		\end{tikzpicture}
		\caption{Original vs Modified Fabric}
	\end{subfigure}
	\hfill
	\begin{subfigure}[b]{0.48\textwidth}
		\centering
		\begin{tikzpicture}
			\begin{axis}[
				xlabel={Number of Transactions},
				ylabel={Avg Response Time (ms)},
				legend style={at={(0.5,1.1)}, anchor=south, font=\small},
				legend columns=2,
				xtick={1000,2000,3000,4000,5000},
				ymajorgrids=true,
				grid=major,
				width=\textwidth,
				height=0.7\textwidth
				]
				\addplot coordinates {(1000,164.1) (2000,157.1) (3000,148.8) (4000,142.2) (5000,150.1)};
				\addplot coordinates {(1000,98.4) (2000,92.4) (3000,96.3) (4000,94.9) (5000,95.8)};
				\addplot coordinates {(1000,110.3) (2000,117.5) (3000,109.1) (4000,107.8) (5000,122)};
				\addplot coordinates {(1000,110.9) (2000,112.2) (3000,121.7) (4000,118.1) (5000,131.4)};
				\legend{Original, Modified, Modified-2T, Modified-4T}
			\end{axis}
		\end{tikzpicture}
		\caption{Parallelism Comparison}
	\end{subfigure}
    \caption{Aborted Transactions Results}
    \label{fig:aborted}
\vspace{-2em}
\end{figure}

\noindent
\textbf{Average Response Time for Committed and Aborted Transactions:} \\
The Modified Fabric consistently reduced the average response time for both committed and aborted transactions compared to the Original Fabric across all transaction volumes (see \Figref{committed} and \Figref{aborted}). At 1000 transactions, the average response time for committed transactions dropped from 215 ms to 132 ms, while for aborted transactions, it fell from 165 ms to 98.4 ms. Notably, the 4-threaded variant sometimes matched or slightly outperformed the dynamic-threaded Modified Fabric in handling committed transactions, suggesting that fixed parallelism can occasionally align well with the underlying DAG structure. However, at higher volumes (e.g., 5000 transactions), the dynamic approach maintained lower and more stable response times, especially for committed transactions. In contrast, the fixed-threaded setups—particularly the 4-threaded configuration—showed elevated response times for aborted transactions (e.g., 131.4 ms at 5000 transactions), likely due to increased thread contention and overhead. These results emphasize the adaptability and responsiveness of dynamic threading under varying workloads and transaction outcomes.

\noindent
\textbf{Key Insights from Experiments:}
The following observations are derived from performance evaluations conducted on the \textbf{Voting smart contract}. Each experiment examines a specific aspect of system performance under various execution strategies, comparing the baseline Original Fabric with optimized versions. All performance metrics reflect either throughput (transactions per second) or latency (in milliseconds).


The Modified Fabric with Dynamic Threading consistently outperforms the baseline across all experiments. It adapts to DAG width dynamically, providing better utilization of parallelism without overwhelming system resources. Fixed-thread approaches (2T/4T per level) can offer benefits but are sensitive to DAG structure and thread scheduling efficiency. These results highlight the value of adaptive concurrency in smart contract execution models.

\section{Related Work}
\label{sec:relwork}

Recent research has proposed various improvements for permissioned blockchains. Androulaki et al.~\cite{androulaki+:2018:eurosys} introduced Fabric’s execute-order-validate architecture. DAG-based approaches have emerged as effective alternatives to improve concurrency \cite{wang+:2024:acmcs}. For instance, BlockPilot \cite{zhao+:2023:icpp} and RT-DAG \cite{zhang+:2024:ieeetpds} enable concurrent execution using dependency-aware scheduling. Constructing order restricted DAG for execution and sharing the DAG for validation stage are explored for generic blockchains framework \cite{Manaswini+:2023:europar,Anjana:OptSC:PDP:2019}. Coloring the contructed DAG to rearrange the DAG levels to get optimized performance is introduced throug Batch-schedule-Execute \cite{yaron+:2024:BSE:SRDS} recently.

Particularly for Hyperledger Fabric, Fabric++ \cite{nasir+:2022:nsdi} and FastFabric \cite{gorenflo+:2020:usenix} enhance throughput via endorsement and validation optimizations, though they retain sequential commitment. Public blockchain systems like Omniledger \cite{kokoris+:2020:sosp}, RapidChain \cite{zamani+:2018:ccs}, and NutBolt \cite{zheng+:2018:osdi} introduce parallel models with distinct trust and consensus assumptions.


\noindent
Unlike prior approaches that require significant architectural changes or new consensus layers, our method preserves full compatibility with Fabric’s chaincode, client APIs, and consensus modules. We implement our modifications on Fabric v2.5 and validate them through experiments on a local Docker-based testbed with varying transaction volumes and dependency levels. Results show up to $40\%$ improvement in throughput and significant reduction in average response time for both committed and aborted transactions.

%% file: conc.tex
Hyperledger Fabric has emerged as a prevalent permissioned blockchain framework; however, its transaction processing mechanism lacks concurrency-awareness, limiting performance under load. This paper introduced a dependency-aware transaction execution scheme for Fabric that identifies inter-transaction dependencies and uses DAG-based block structures to enable parallelism.

\noindent
Our proposed enhancements include: dependency flagging at the endorsement phase, dependency-preserving block construction at the ordering service, and DAG-guided execution at the committer. The implementation extends the existing Fabric codebase with minimal intrusion, ensuring compatibility and maintainability. Experimental evaluations demonstrate that the modified Fabric significantly outperforms the standard implementation across key metrics. Notably, throughput increases by up to $25\%$, response times are consistently lower, and commit rates remain higher under varying transaction loads and dependency ratios. These improvements are especially evident in scenarios with high contention, where the original Fabric struggles to maintain performance.

In future work, we plan to integrate dynamic dependency detection using static code analysis or machine learning to remove the need for client-declared read/write sets. Additionally, incorporating fault-tolerant and Byzantine-resilient endorser leader election and DAG commit logic would make the solution more robust in distributed deployments. Further exploration of integration with Fabric private data collections and channels could enhance support for confidential multi-party workflows.

%% file: append.tex
\begin{center}
\textbf{Appendix Table of Contents}
\end{center}

\begin{enumerate}
  \item \textbf{Set-up Summary} \dotfill \pageref{apn:setup}
  \item \textbf{Security and Correctness} \dotfill \pageref{apn:sec}
  \item \textbf{Asset-Transfer Contract Results} \dotfill \pageref{apn:asset}
  \item \textbf{Wallet Contract Results} \dotfill \pageref{apn:wallet}
\end{enumerate}

\subsection{Set-up Summary}
\label{apn:setup}

\noindent
This environment was used consistently across all experiments to ensure a fair comparison of the original and modified Fabric behavior. Several key components of Hyperledger Fabric were modified to support the evaluation of dependency-aware transaction processing. The modifications are detailed in \Tabref{components}. We have introduced the concept of leader endorser to coordinate the dependency for all the transactions received by the endorsing peers. To simulate realistic and diverse workloads, the experimental setup varied several parameters, detailed in \Tabref{parameters}. Each block contained between 1000 and 5000 transactions, with the proportion of transactions having dependencies ranging from 0\% (no inter-transaction dependencies) to 90\% (highly dependent workloads).

\begin{table}[h]
	\centering
	\begin{tabular}{|l|p{10cm}|}
		\hline
		\textbf{Component} & \textbf{Modification} \\
		\hline
		Leader Endorser & Added logic to detect key conflicts and tag transactions with dependency flags \\
		Committer & Implemented DAG constructor and thread pool-based executor for level-wise validation \\
		Protobuf Files & Extended block metadata to include transaction dependency markers \\
		\hline
	\end{tabular}
	\caption{Key Modifications in Hyperledger Fabric}
    \label{tab:components}
\end{table}
\vspace{-3em}
\begin{table}[h]
	\centering
	\begin{tabular}{|l|l|}
		\hline
		\textbf{Parameter} & \textbf{Value / Range} \\
		\hline
		Transactions per block & 1000 -- 5000 \\
		Dependency ratio & 0\% -- 90\% \\
		Retry logic & Disabled \\
		Threads per committer & 2, 4 and dynamic threads in DAG levels \\
		Transaction logic & Mix of read/write operations on shared key-value space \\
		\hline
	\end{tabular}
	\caption{Workload Simulation Parameters}
    \label{tab:parameters}
\end{table}

\subsection{Security and Correctness}
\label{apn:sec}

Any architectural modification to a permissioned blockchain must maintain the integrity, determinism, and fault tolerance of the original system. Our proposed enhancements to Hyperledger Fabric were designed with these properties in mind, ensuring that the changes do not compromise the security guarantees already provided by the platform.

\noindent
\textbf{Data Consistency and Validation:} Fabric maintains correctness through version checks and endorsement policies. Our model preserves this by not altering the semantics of transaction simulation, endorsement, or validation. Each transaction still undergoes the same chaincode-based simulation at endorsers. The DAG formation and utilisation remove the need for version check without altering the logic of the version check (in case that would have been in place). For example, let's suppose we have 2 endorsed transactions which update the same asset in the world state. In the case of original logic, the first transaction will succeed, but the second transaction will be rejected. This is because the version of the asset will be updated after the completion of the first transaction. The endorsed version in the second transaction would not match the new version of the asset now. In our updated code base, the second transaction would be checked as per the chain code. Passing the check would mean that the transaction would not cause any conflicts and could be executed without any issues. Hence, the proposed architecture reduces transaction rejections without accepting any conflicting transactions. The dependency flagging added during endorsement is the metadata and is used only for scheduling. It does not change the transaction’s logic or state transitions.

\noindent
Hence, incorrect or conflicting transactions are filtered out before updating the world state. This ensures that the integrity of the ledger is preserved, even when multiple transactions are executed in parallel.

\noindent
\textbf{Determinism via DAG Execution:} The DAG structure enforces a partial order among dependent transactions. The use of topological sorting ensures that parent transactions are always processed before their children. This guarantees that the final ledger state is deterministic and consistent across all committing peers, regardless of thread scheduling or parallelism.

\noindent
The DAG is generated from explicit flags and known dependencies, and therefore remains the same across all peers. No randomness or non-deterministic ordering is introduced in the execution pipeline.

\noindent
\textbf{Concurrency Control:} Only transactions that are marked as independent and conflict-free are allowed to execute in parallel. Any transaction with dependencies is delayed until its parent transaction(s) complete. This avoids race conditions and ensures that no transaction is executed in an invalid state.

\noindent
All shared structures used during DAG scheduling and commit (e.g., state buffers, world state locks) are synchronized using thread-safe mechanisms, ensuring that data races or interleaved commits do not corrupt state.

\noindent
\textbf{Endorsement Validity and Expiry:} To prevent stale or reused endorsements from affecting correctness, each endorsed transaction includes a validity timeout. If a transaction does not get committed within this time window, its endorsement is invalidated and the transaction is dropped. This mechanism prevents accumulation of expired or invalid dependency links and keeps the endorsement context clean.


\noindent
\textbf{Security Summary:} In summary, our changes do not bypass any of the validation or consensus stages in Fabric. They preserve:
\begin{itemize}
	\item \textbf{Ledger correctness:} All committed transactions are validated and applied in a version-consistent order.
	\item \textbf{Execution determinism:} DAG execution guarantees identical ordering and results across peers.
	\item \textbf{Concurrency safety:} All parallel execution is strictly controlled using dependency-aware scheduling.
	\item \textbf{Peer consistency:} All peers construct the same DAG and validate transactions independently.
\end{itemize}

\noindent
Thus, the proposed system maintains the security, correctness, and trust guarantees of Hyperledger Fabric while improving its performance and scalability.

\subsection{Asset-Transfer Contract Results}
\label{apn:asset}
This section summarizes the experimental findings for the Asset-Transfer smart contract, evaluating performance across different execution strategies: the baseline Original Fabric, the Modified Fabric with Dynamic Threads, and fixed-threaded variants (2 threads and 4 threads per DAG level). The results are grouped by transaction volume and dependency ratio to assess performance under varying conditions.

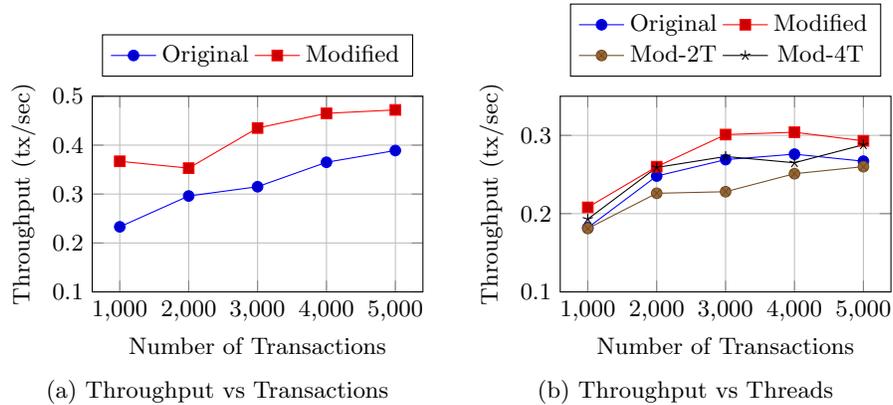
\begin{figure}[htbp]
	\centering
	
	\begin{subfigure}{0.49\textwidth}
		\centering
		\begin{tikzpicture}
			\begin{axis}[
				width=\textwidth,
				height=0.7\textwidth,
				xlabel={Number of Transactions},
				ylabel={Throughput (tx/sec)},
				xtick=data,
				ymin=0.1,
				ymax=0.5,
				grid=major,
				legend style={at={(0.5,1.1)}, anchor=south, legend columns=2},
				]
				\addplot coordinates {(1000,0.233) (2000,0.296) (3000,0.315) (4000,0.365) (5000,0.389)};
				\addplot coordinates {(1000,0.367) (2000,0.353) (3000,0.435) (4000,0.465) (5000,0.472)};
				\legend{Original, Modified}
			\end{axis}
		\end{tikzpicture}
		\caption{Throughput vs Transactions}
	\end{subfigure}
	\hfill
	\begin{subfigure}{0.49\textwidth}
		\centering
		\begin{tikzpicture}
			\begin{axis}[
				width=\textwidth,
				height=0.7\textwidth,
				xlabel={Number of Transactions},
				ylabel={Throughput (tx/sec)},
				xtick=data,
				ymin=0.1,
				ymax=0.35,
				grid=major,
				legend style={at={(0.5,1.1)}, anchor=south, legend columns=2},
				]
				\addplot coordinates {(1000,0.182) (2000,0.248) (3000,0.269) (4000,0.276) (5000,0.267)};
				\addplot coordinates {(1000,0.208) (2000,0.260) (3000,0.301) (4000,0.304) (5000,0.293)};
				\addplot coordinates {(1000,0.181) (2000,0.226) (3000,0.228) (4000,0.251) (5000,0.260)};
				\addplot coordinates {(1000,0.193) (2000,0.259) (3000,0.273) (4000,0.265) (5000,0.288)};
				\legend{Original, Modified, Mod-2T, Mod-4T}
			\end{axis}
		\end{tikzpicture}
		\caption{Throughput vs Threads}
	\end{subfigure}
	
	\caption{Throughput Analysis of Asset Transfer Contract on Fabric Variants}
\end{figure}

\textbf{Experiment 1: Throughput vs Number of Transactions} \\
The Modified Fabric demonstrated consistent throughput improvement over the Original Fabric. For example, at 5000 transactions, throughput increased from 0.389 tx/sec to 0.472 tx/sec. The fixed-threaded versions (2T and 4T) showed moderate performance improvements but were generally less adaptive. The 4-threaded variant showed better throughput than the 2-threaded one, but the Modified Fabric with dynamic thread allocation outperformed both by adjusting concurrency per DAG level based on available cores and transaction density.

\begin{figure}[htbp]
    \centering

    \begin{subfigure}{0.49\textwidth}
        \centering
        \begin{tikzpicture}
            \begin{axis}[
                width=\textwidth,
                height=0.7\textwidth,
                xlabel={Txns},
                ylabel={Avg RT (ms)},
                xtick=data,
                grid=major,
                legend style={at={(0.5,1.1)}, anchor=south, legend columns=2},
            ]
                \addplot coordinates {(1000,295) (2000,243) (3000,258) (4000,228) (5000,219)};
                \addplot coordinates {(1000,168) (2000,196) (3000,167) (4000,166) (5000,173)};
                \legend{Original, Modified}
            \end{axis}
        \end{tikzpicture}
        \caption{All Txns}
    \end{subfigure}
    \hfill
    \begin{subfigure}{0.49\textwidth}
        \centering
        \begin{tikzpicture}
            \begin{axis}[
                width=\textwidth,
                height=0.7\textwidth,
                xlabel={Txns},
                ylabel={Avg RT (ms)},
                xtick=data,
                grid=major,
                legend style={at={(0.5,1.1)}, anchor=south, legend columns=2},
            ]
                \addplot coordinates {(1000,446.3) (2000,343.3) (3000,332.7) (4000,334) (5000,337.7)};
                \addplot coordinates {(1000,290) (2000,289) (3000,295.3) (4000,301) (5000,307.3)};
                \addplot coordinates {(1000,365.5) (2000,350.7) (3000,385.8) (4000,352.8) (5000,349.5)};
                \addplot coordinates {(1000,420.5) (2000,289.3) (3000,307) (4000,339) (5000,310.7)};
                \legend{Orig, Mod, 2T, 4T}
            \end{axis}
        \end{tikzpicture}
        \caption{All Txns - Threads}
    \end{subfigure}

    \caption{Average Response Time - All Transactions}
\end{figure}
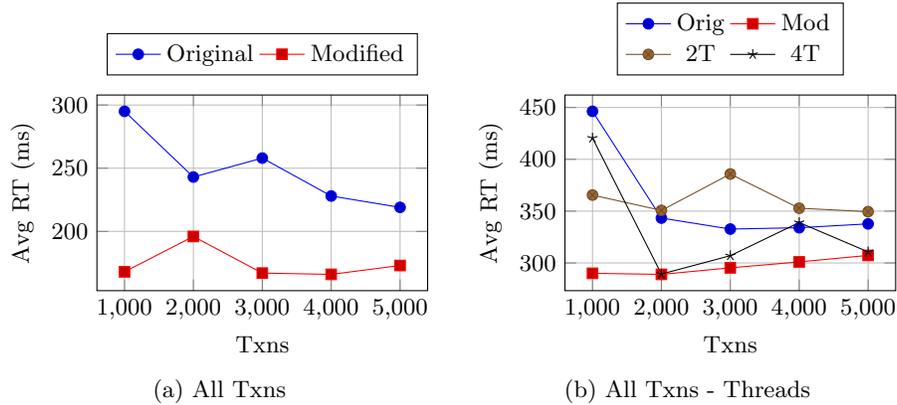

\noindent
\textbf{Experiment 2: Average Response Time vs Number of Transactions} \\
In terms of average response time across all transactions, the Modified Fabric offered lower latency in most scenarios. For instance, at 1000 transactions, latency dropped from 295 ms to 168 ms. However, both 2-threaded and 4-threaded versions showed inconsistent behavior: sometimes approaching the Modified Fabric's performance and other times suffering from overhead, particularly in the 4T configuration due to excessive context switching and synchronization.

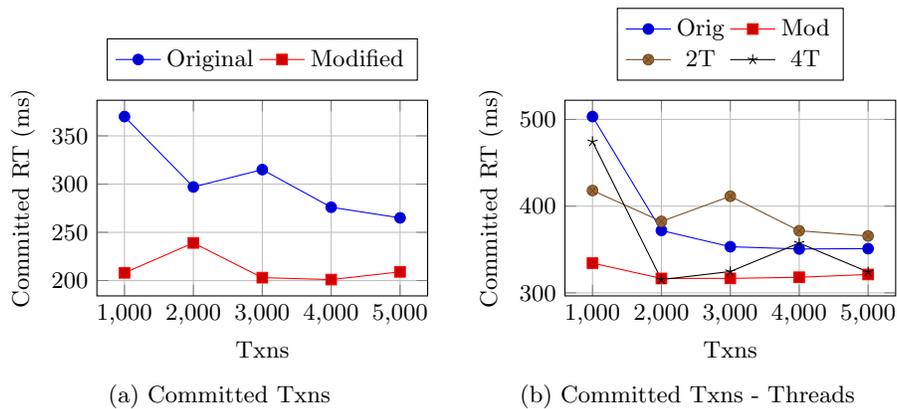
\begin{figure}[htbp]
    \centering

    \begin{subfigure}{0.49\textwidth}
        \centering
        \begin{tikzpicture}
            \begin{axis}[
                width=\textwidth,
                height=0.7\textwidth,
                xlabel={Txns},
                ylabel={Committed RT (ms)},
                xtick=data,
                grid=major,
                legend style={at={(0.5,1.1)}, anchor=south, legend columns=2},
            ]
                \addplot coordinates {(1000,370) (2000,297) (3000,315) (4000,276) (5000,265)};
                \addplot coordinates {(1000,208) (2000,239) (3000,203) (4000,201) (5000,209)};
                \legend{Original, Modified}
            \end{axis}
        \end{tikzpicture}
        \caption{Committed Txns}
    \end{subfigure}
    \hfill
    \begin{subfigure}{0.49\textwidth}
        \centering
        \begin{tikzpicture}
            \begin{axis}[
                width=\textwidth,
                height=0.7\textwidth,
                xlabel={Txns},
                ylabel={Committed RT (ms)},
                xtick=data,
                grid=major,
                legend style={at={(0.5,1.1)}, anchor=south, legend columns=2},
            ]
                \addplot coordinates {(1000,503.3) (2000,371.9) (3000,353.2) (4000,350.6) (5000,351)};
                \addplot coordinates {(1000,334.3) (2000,316.6) (3000,316.7) (4000,318) (5000,321.2)};
                \addplot coordinates {(1000,418) (2000,382.4) (3000,411.4) (4000,371.6) (5000,365.6)};
                \addplot coordinates {(1000,474.4) (2000,315.3) (3000,324.6) (4000,357.4) (5000,324.1)};
                \legend{Orig, Mod, 2T, 4T}
            \end{axis}
        \end{tikzpicture}
        \caption{Committed Txns - Threads}
    \end{subfigure}

    \caption{Average Response Time - Committed Transactions}
\end{figure}

\noindent
\textbf{Experiment 2 (Committed Transactions):} \\
The response time for committed transactions also improved significantly with the Modified Fabric. For instance, at 1000 transactions, latency dropped from 370 ms (Original) to 208 ms (Modified). While the 2T and 4T versions yielded occasional benefits (e.g., 315.3 ms for 4T at 2000 transactions), they frequently underperformed compared to the adaptive Modified Fabric, which maintained a consistent edge.

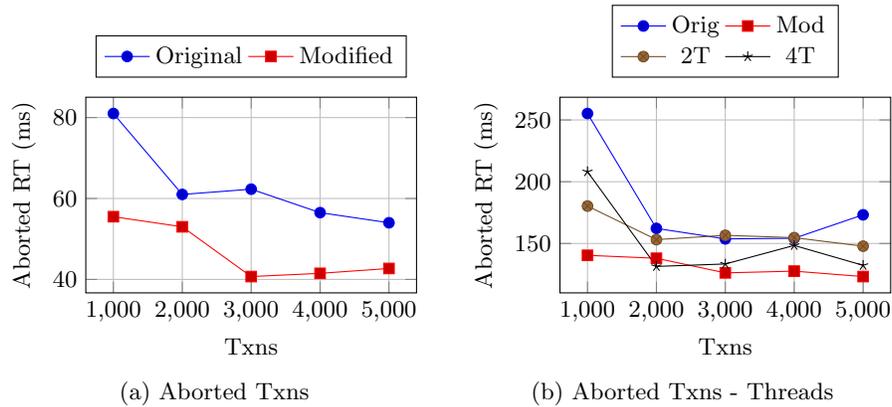
\begin{figure}[htbp]
    \centering

    \begin{subfigure}{0.49\textwidth}
        \centering
        \begin{tikzpicture}
            \begin{axis}[
                width=\textwidth,
                height=0.7\textwidth,
                xlabel={Txns},
                ylabel={Aborted RT (ms)},
                xtick=data,
                grid=major,
                legend style={at={(0.5,1.1)}, anchor=south, legend columns=2},
            ]
                \addplot coordinates {(1000,81) (2000,61) (3000,62.3) (4000,56.5) (5000,54)};
                \addplot coordinates {(1000,55.5) (2000,53) (3000,40.7) (4000,41.5) (5000,42.7)};
                \legend{Original, Modified}
            \end{axis}
        \end{tikzpicture}
        \caption{Aborted Txns}
    \end{subfigure}
    \hfill
    \begin{subfigure}{0.49\textwidth}
        \centering
        \begin{tikzpicture}
            \begin{axis}[
                width=\textwidth,
                height=0.7\textwidth,
                xlabel={Txns},
                ylabel={Aborted RT (ms)},
                xtick=data,
                grid=major,
                legend style={at={(0.5,1.1)}, anchor=south, legend columns=2},
            ]
                \addplot coordinates {(1000,255.2) (2000,162.2) (3000,153.7) (4000,154.2) (5000,173.2)};
                \addplot coordinates {(1000,140.4) (2000,138) (3000,126.2) (4000,127.6) (5000,123.2)};
                \addplot coordinates {(1000,180.3) (2000,153.1) (3000,156.6) (4000,154.7) (5000,147.8)};
                \addplot coordinates {(1000,208.1) (2000,131.4) (3000,133.4) (4000,148.4) (5000,132.3)};
                \legend{Orig, Mod, 2T, 4T}
            \end{axis}
        \end{tikzpicture}
        \caption{Aborted Txns - Threads}
    \end{subfigure}

    \caption{Average Response Time - Aborted Transactions}
\end{figure}

\noindent
\textbf{Experiment 2 (Aborted Transactions):} \\
The Modified Fabric also reduced the response time for aborted transactions. At 1000 transactions, the average latency dropped from 81 ms to 55.5 ms. The fixed-threaded approaches lagged behind in this category, with the 4-threaded variant recording 208.1 ms at the same transaction level, indicating inefficiencies in abort handling with rigid thread allocation.

\begin{figure}[htbp]
	\centering
	\caption{Experiment 3: Dependency Ratio vs Average Response Time}

   \begin{subfigure}{0.49\textwidth}
        \centering
        \begin{tikzpicture}
            \begin{axis}[
                width=\textwidth,
                height=0.7\textwidth,
                xlabel={Dependency Ratio},
                ylabel={Aborted RT (ms)},
                xtick=data,
                grid=major,
                legend style={at={(0.5,1.1)}, anchor=south,legend columns=2},
            ]
                \addplot coordinates {(0.1,229) (0.2,200) (0.3,314) (0.4,231) (0.5,200) (0.6,226) (0.7,322) (0.8,289) (0.9,370)};
                \addplot coordinates {(0.1,190) (0.2,193) (0.3,168) (0.4,156) (0.5,163) (0.6,207) (0.7,160) (0.8,210) (0.9,218)};
                \legend{Original, Modified}
            \end{axis}
        \end{tikzpicture}
        \caption{Aborted Txns}
    \end{subfigure}
    \hfill
    \begin{subfigure}{0.49\textwidth}
        \centering
        \begin{tikzpicture}
            \begin{axis}[
                width=\textwidth,
                height=0.7\textwidth,
                xlabel={Dependency Ratio},
                ylabel={Aborted RT (ms)},
                xtick=data,
                grid=major,
                legend style={at={(0.5,1.1)}, anchor=south, legend columns=2},
            ]
                \addplot coordinates {(0.1,212) (0.2,195) (0.3,172) (0.4,181) (0.5,255) (0.6,180)};
                \addplot coordinates {(0.1,158) (0.2,175) (0.3,195) (0.4,150) (0.5,140) (0.6,145)};
                \addplot coordinates {(0.1,220) (0.2,200) (0.3,197) (0.4,236) (0.5,181) (0.6,186)};
                \addplot coordinates {(0.1,175) (0.2,150) (0.3,155) (0.4,145) (0.5,210) (0.6,145)};
                \legend{Orig, Mod, 2T, 4T}
            \end{axis}
        \end{tikzpicture}
        \caption{Aborted Txns - Threads}
    \end{subfigure}

    \caption{Avg Response Time vs Dependency Ratio for Aborted Transactions}
	
\end{figure}

\vspace{1em}
\noindent
\textbf{Experiment 3: Response Time vs Dependency Ratio} \\
As inter-transaction dependencies increased, the Original Fabric showed a dramatic increase in latency (e.g., 642 ms at dependency ratio 0.3). In contrast, the Modified Fabric was significantly more stable, maintaining response times in the 290–375 ms range across all ratios. The 2T and 4T variants showed mixed outcomes—performing well at lower dependency ratios but struggling at higher ones, likely due to rigid thread scheduling unable to adapt to dependency-heavy DAG structures.

\vspace{1em}
\noindent
\textbf{Experiment 3 (Committed Transactions):} \\
The Modified Fabric consistently outperformed the baseline in response time for committed transactions under all dependency ratios. At 0.3, the Original took 707 ms versus 353 ms for Modified. Again, fixed-threaded approaches sometimes matched the Modified Fabric (e.g., 344 ms for 4T at 0.2) but failed to maintain consistent gains due to their static nature.

\vspace{1em}
\noindent
\textbf{Experiment 3 (Aborted Transactions):} \\
Aborted transaction latency also favored the Modified Fabric. For example, at dependency ratio 0.3, the Modified Fabric recorded 168 ms, while the Original showed 314 ms. The 2T and 4T variants occasionally yielded comparable results (e.g., 145 ms for 4T at 0.6), but performance varied based on workload characteristics, again showing the superior adaptability of the dynamic threading model.


\subsection{Wallet Contract Results}
\label{apn:wallet}

This section of appendix presents a detailed performance evaluation of the \textbf{Wallet Contract} deployed on Hyperledger Fabric. The results compare the \textbf{Original Fabric} against an optimized \textbf{Modified Fabric} version, including multi-threaded variants using \textbf{2-thread} and \textbf{4-thread} configurations. 

\begin{figure}[htbp]
    \centering

    \begin{subfigure}{0.49\textwidth}
        \centering
        \begin{tikzpicture}
            \begin{axis}[
                width=\textwidth,
                height=0.7\textwidth,
                xlabel={Transactions},
                ylabel={Throughput (tx/sec)},
                ymin=0.12, ymax=0.23,
                xtick=data,
                grid=major,
                legend style={at={(0.5,1.1)}, anchor=south, legend columns=2},
            ]
                \addplot coordinates {(1000,0.148) (2000,0.163) (3000,0.187) (4000,0.184) (5000,0.180)};
                \addplot coordinates {(1000,0.165) (2000,0.203) (3000,0.200) (4000,0.217) (5000,0.203)};
                \legend{Original, Modified}
            \end{axis}
        \end{tikzpicture}
        \caption{Original vs Modified}
    \end{subfigure}
    \hfill
    \begin{subfigure}{0.49\textwidth}
        \centering
        \begin{tikzpicture}
            \begin{axis}[
                width=\textwidth,
                height=0.7\textwidth,
                xlabel={Transactions},
                ylabel={Throughput (tx/sec)},
                ymin=0.12, ymax=0.23,
                xtick=data,
                grid=major,
                legend style={at={(0.5,1.1)}, anchor=south, legend columns=2},
                title={Thread Scaling}
            ]
                \addplot coordinates {(1000,0.136) (2000,0.161) (3000,0.168) (4000,0.164) (5000,0.154)};
                \addplot coordinates {(1000,0.165) (2000,0.196) (3000,0.177) (4000,0.198) (5000,0.191)};
                \addplot coordinates {(1000,0.133) (2000,0.161) (3000,0.181) (4000,0.190) (5000,0.167)};
                \addplot coordinates {(1000,0.137) (2000,0.160) (3000,0.183) (4000,0.158) (5000,0.149)};
                \legend{Orig, Mod, 2T, 4T}
            \end{axis}
        \end{tikzpicture}
        \caption{Modified Fabric with Threads}
    \end{subfigure}

    \caption{Wallet Contract - Experiment 1: Throughput vs Number of Transactions}
\end{figure}
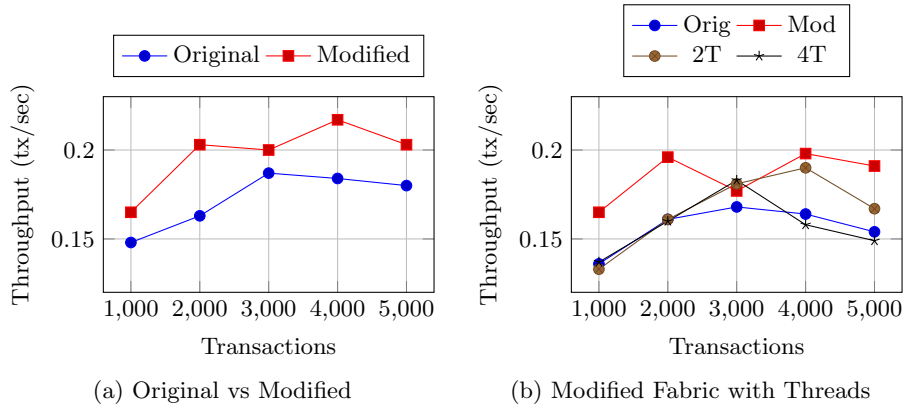
\noindent
\textbf{Experiment 1}: Throughput vs Number of Transactions
analyzes throughput (in transactions per second) as the number of submitted transactions scales from 1,000 to 5,000.

\begin{itemize}
	\item \textbf{Original Fabric} achieved a maximum throughput of \textbf{0.187 tx/sec}.
	\item \textbf{Modified Fabric} improved performance across all cases, peaking at \textbf{0.217 tx/sec}.
	\item The multi-threaded configurations did not consistently improve throughput; in some cases, throughput degraded due to thread contention or synchronization overhead.
\end{itemize}
\begin{figure}[htbp]
	\centering
	\caption{Wallet Contract - Experiment 2: Avg Response Time for All Transactions}
	
	\begin{subfigure}{0.40\textwidth}
		\centering
		\begin{tikzpicture}
			\begin{axis}[
				title={Original vs Modified},
				xlabel={Transactions}, ylabel={Avg RT (ms)},
				width=\textwidth, height=4.2cm,
				ymin=60, ymax=110,
				tick label style={font=\small},
				label style={font=\small},
				title style={font=\small},
				legend style={font=\tiny, at={(0.5,-0.3)}, anchor=north}
			]
				\addplot coordinates {(1000,88.5) (2000,84.8) (3000,81.6) (4000,80.2) (5000,87.3)};
				\addplot coordinates {(1000,66.4) (2000,67) (3000,74.8) (4000,71) (5000,75.8)};
				\legend{Original, Modified}
			\end{axis}
		\end{tikzpicture}
		\caption{All Txns}
	\end{subfigure}
	\hfill
	\begin{subfigure}{0.40\textwidth}
		\centering
		\begin{tikzpicture}
			\begin{axis}[
				title={Thread Scaling},
				xlabel={Transactions}, ylabel={Avg RT (ms)},
				width=\textwidth, height=4.2cm,
				ymin=60, ymax=110,
				tick label style={font=\small},
				label style={font=\small},
				title style={font=\small},
				legend style={font=\tiny, at={(0.5,-0.35)}, anchor=north, legend columns=2}
			]
				\addplot coordinates {(1000,90) (2000,88) (3000,86.3) (4000,90) (5000,99)};
				\addplot coordinates {(1000,68) (2000,69.4) (3000,85.3) (4000,74.6) (5000,83.2)};
				\addplot coordinates {(1000,103.6) (2000,87.8) (3000,82.5) (4000,76.2) (5000,90.3)};
				\addplot coordinates {(1000,92.3) (2000,86) (3000,81) (4000,94) (5000,103.1)};
				\legend{Orig, Mod, 2T, 4T}
			\end{axis}
		\end{tikzpicture}
		\caption{All Txns - Threads}
	\end{subfigure}
\end{figure}
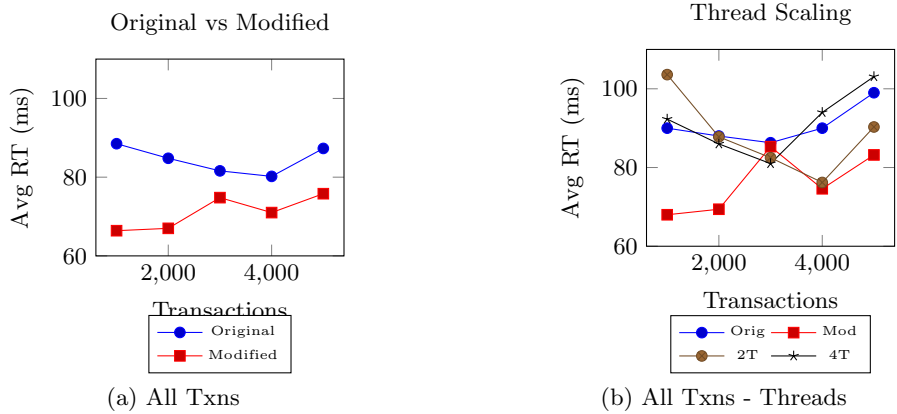

\noindent
\textbf{Experiment 2:} Average Response Time vs Number of Transactions measures the average response time (in milliseconds) under increasing transaction volume, separated by:

\textbf{All Transactions:}
\begin{itemize}
	\item The Modified Fabric reduced average response time significantly (e.g., \textbf{66.4 ms} vs \textbf{88.5 ms} at 1,000 transactions).
	\item The 2-thread configuration often provided a slight benefit; the 4-thread version showed inconsistent gains.
\end{itemize}

\textbf{Committed Transactions}
\begin{itemize}
	\item The Modified Fabric consistently outperformed the Original, with improvements up to \textbf{20\%}.
	\item The 2-thread configuration provided better latency under higher load compared to 4-thread, which occasionally suffered from synchronization delays.
\end{itemize}

\begin{figure}[htbp]
	\centering
	\caption{Wallet Contract - Experiment 2: Avg Response Time for Committed Transactions}
	
	\begin{subfigure}{0.40\textwidth}
		\centering
		\begin{tikzpicture}
			\begin{axis}[
				title={Original vs Modified},
				xlabel={Transactions}, ylabel={Committed RT (ms)},
				width=\textwidth, height=4.2cm,
				ymin=70, ymax=120,
				tick label style={font=\small},
				label style={font=\small},
				title style={font=\small},
				legend style={font=\tiny, at={(0.5,-0.3)}, anchor=north}
			]
				\addplot coordinates {(1000,93.5) (2000,89.5) (3000,86.3) (4000,84.5) (5000,93)};
				\addplot coordinates {(1000,72.2) (2000,71.1) (3000,80.8) (4000,75.5) (5000,81.1)};
				\legend{Original, Modified}
			\end{axis}
		\end{tikzpicture}
		\caption{Committed Txns}
	\end{subfigure}
	\hfill
	\begin{subfigure}{0.40\textwidth}
		\centering
		\begin{tikzpicture}
			\begin{axis}[
				title={Thread Scaling},
				xlabel={Transactions}, ylabel={Committed RT (ms)},
				width=\textwidth, height=4.2cm,
				ymin=70, ymax=120,
				tick label style={font=\small},
				label style={font=\small},
				title style={font=\small},
				legend style={font=\tiny, at={(0.5,-0.35)}, anchor=north, legend columns=2}
			]
				\addplot coordinates {(1000,95.5) (2000,92.8) (3000,91.9) (4000,94.8) (5000,105.3)};
				\addplot coordinates {(1000,73.5) (2000,74) (3000,92.5) (4000,80.2) (5000,89.7)};
				\addplot coordinates {(1000,112) (2000,93.1) (3000,86.3) (4000,80.4) (5000,96.2)};
				\addplot coordinates {(1000,98.5) (2000,91.7) (3000,85.6) (4000,101.4) (5000,110.7)};
				\legend{Orig, Mod, 2T, 4T}
			\end{axis}
		\end{tikzpicture}
		\caption{Committed Txns - Threads}
	\end{subfigure}
\end{figure}
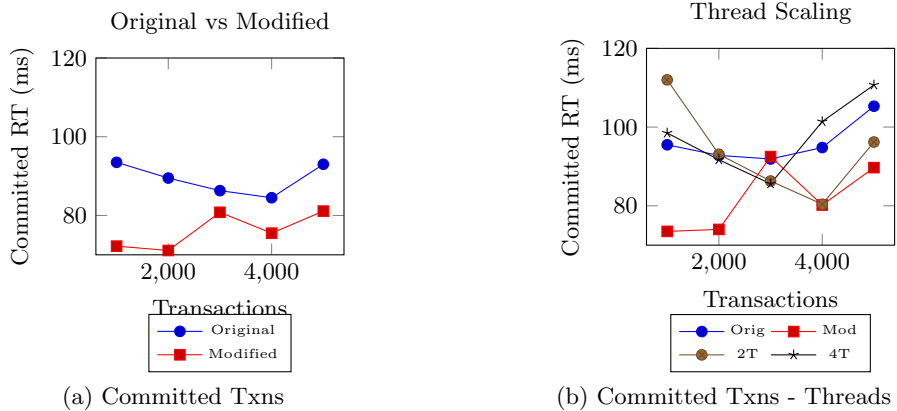

\textbf{Aborted Transactions}
\begin{itemize}
	\item A marked improvement was seen in the Modified Fabric, reducing response time from \textbf{59.4 ms} to as low as \textbf{33.8 ms}.
	\item Multi-threaded versions did not consistently improve performance for aborted transactions and occasionally increased variability.
\end{itemize}

\begin{figure}[htbp]
	\centering
	\caption{Wallet Contract - Experiment 2: Avg Response Time for Aborted Transactions}
	
	\begin{subfigure}{0.40\textwidth}
		\centering
		\begin{tikzpicture}
			\begin{axis}[
				title={Original vs Modified},
				xlabel={Transactions}, ylabel={Aborted RT (ms)},
				width=\textwidth, height=4.2cm,
				ymin=30, ymax=65,
				tick label style={font=\small},
				label style={font=\small},
				title style={font=\small},
				legend style={font=\tiny, at={(0.5,-0.3)}, anchor=north}
			]
				\addplot coordinates {(1000,59.4) (2000,53.5) (3000,50.1) (4000,48.1) (5000,48)};
				\addplot coordinates {(1000,36.5) (2000,35.1) (3000,33.8) (4000,34.8) (5000,35.3)};
				\legend{Original, Modified}
			\end{axis}
		\end{tikzpicture}
		\caption{Aborted Txns}
	\end{subfigure}
	\hfill
	\begin{subfigure}{0.40\textwidth}
		\centering
		\begin{tikzpicture}
			\begin{axis}[
				title={Thread Scaling},
				xlabel={Transactions}, ylabel={Aborted RT (ms)},
				width=\textwidth, height=4.2cm,
				ymin=30, ymax=65,
				tick label style={font=\small},
				label style={font=\small},
				title style={font=\small},
				legend style={font=\tiny, at={(0.5,-0.35)}, anchor=north, legend columns=2}
			]
				\addplot coordinates {(1000,59.5) (2000,54.2) (3000,50.5) (4000,50.4) (5000,49.2)};
				\addplot coordinates {(1000,36.9) (2000,38.1) (3000,35.1) (4000,35.8) (5000,36.9)};
				\addplot coordinates {(1000,60.3) (2000,52.5) (3000,49.4) (4000,47.7) (5000,47.1)};
				\addplot coordinates {(1000,54.3) (2000,54.2) (3000,46.7) (4000,48) (5000,45.9)};
				\legend{Orig, Mod, 2T, 4T}
			\end{axis}
		\end{tikzpicture}
		\caption{Aborted Txns - Threads}
	\end{subfigure}
\end{figure}

\begin{figure}[htbp]
	\centering
	\caption{Wallet Contract - Experiment 3: Average Response Time vs Dependency Ratio}

 \begin{subfigure}{0.40\textwidth}
        \centering
        \begin{tikzpicture}
            \begin{axis}[
                title={Original vs Modified},
                xlabel={Dependency Ratio}, ylabel={Aborted RT (ms)},
                width=\textwidth, height=4.2cm,
                ymin=30, ymax=75,
                tick label style={font=\small},
                label style={font=\small},
                title style={font=\small},
                legend style={font=\tiny, at={(0.5,-0.3)}, anchor=north}
            ]
                \addplot coordinates {
                    (0.1,65.7) (0.2,62.6) (0.3,60.9) (0.4,60.9) (0.5,59.2) 
                    (0.6,59.8) (0.7,58.5) (0.8,59.4) (0.9,65.2)};
                \addplot coordinates {
                    (0.1,39.8) (0.2,36.4) (0.3,37.3) (0.4,36.6) (0.5,36.5) 
                    (0.6,38.9) (0.7,37.3) (0.8,39.6) (0.9,43.1)};
                \legend{Original, Modified}
            \end{axis}
        \end{tikzpicture}
        \caption{Aborted Txns}
    \end{subfigure}
    \hfill
    \begin{subfigure}{0.40\textwidth}
        \centering
        \begin{tikzpicture}
            \begin{axis}[
                title={Thread Scaling},
                xlabel={Dependency Ratio}, ylabel={Aborted RT (ms)},
                width=\textwidth, height=4.2cm,
                ymin=30, ymax=75,
                tick label style={font=\small},
                label style={font=\small},
                title style={font=\small},
                legend style={font=\tiny, at={(0.5,-0.35)}, anchor=north, legend columns=2}
            ]
                \addplot coordinates {
                    (0.1,68) (0.2,63) (0.3,61) (0.4,62) (0.5,60) 
                    (0.6,62) (0.7,60) (0.8,59) (0.9,66)};
                \addplot coordinates {
                    (0.1,42) (0.2,36.5) (0.3,38.5) (0.4,37) (0.5,36.5) 
                    (0.6,41.5) (0.7,37.5) (0.8,42) (0.9,44.8)};
                \addplot coordinates {
                    (0.1,70) (0.2,64) (0.3,60) (0.4,58.3) (0.5,60.4) 
                    (0.6,58.8) (0.7,57.7) (0.8,61.7) (0.9,72.3)};
                \addplot coordinates {
                    (0.1,65.5) (0.2,60) (0.3,59.5) (0.4,62) (0.5,54.3) 
                    (0.6,57) (0.7,58.5) (0.8,59) (0.9,65.6)};
                \legend{Orig, Mod, 2T, 4T}
            \end{axis}
        \end{tikzpicture}
        \caption{Aborted Txns - Threads}
    \end{subfigure}    
    \caption*{\textbf{(c) Avg Response Time for Aborted Transactions}}
	
\end{figure}

\noindent
\textbf{Experiment 3:} Average Response Time vs Dependency Ratio
evaluates system latency across increasing dependency ratios (0.0 to 0.9), simulating transaction contention and logical dependencies.

\textbf{All Transactions:}
\begin{itemize}
	\item The Modified Fabric consistently outperformed the Original, with response times ranging from \textbf{62.6 ms} to \textbf{73.3 ms}.
	\item Multi-threaded variants yielded mixed results—sometimes improving response time (e.g., at ratio 0.3), but occasionally increasing latency due to scheduling inefficiencies.
\end{itemize}

\textbf{Committed Transactions:}
\begin{itemize}
	\item The Modified Fabric offered smoother response time curves across varying ratios.
	\item The 2-thread and 4-thread variants provided benefits at some ratios (e.g., 0.3 to 0.6), though performance often peaked unpredictably at higher contention levels.
\end{itemize}

\textbf{Aborted Transactions:}
\begin{itemize}
	\item The Modified Fabric demonstrated significant improvements, with response times reducing from over \textbf{65 ms} to as low as \textbf{36 ms}.
	\item Multi-threaded variants exhibited moderate benefits in some cases, but also introduced occasional spikes in response time.
\end{itemize}

